\documentclass[aip,apl,reprint,superscriptaddress]{revtex4-1}

\usepackage[version=3]{mhchem} 
\usepackage{graphicx}
\usepackage{hyperref}

\usepackage{physics}
\usepackage{amsmath}
\usepackage{braket}
\usepackage{amssymb}
\usepackage{multirow}
\usepackage{gensymb}

\begin{document}


\title
[\centerline{APL MATERIALS 1, ****** (2018)}] 
{Perspective: Dielectric and Ferroic Properties of Metal Halide Perovskites}
%

\author{Jacob N. Wilson}
\affiliation{Thomas Young Centre and Department of Materials, Imperial College London, London SW7 2AZ, UK} 
\author{Jarvist M. Frost}
\affiliation{Department of Physics, Imperial College London, London SW7 2AZ, UK} 
\affiliation{Department of Physics, King's College London, London WC2R 2LS, UK}
\author{Suzanne K. Wallace}
\affiliation{Thomas Young Centre and Department of Materials, Imperial College London, London SW7 2AZ, UK} 
\affiliation{Department of Chemistry, University of Bath, Claverton Down, Bath, BA2 7AY, UK}
\author{Aron Walsh}
\email[E-mail: ]{a.walsh@imperial.ac.uk}
\affiliation{Thomas Young Centre and Department of Materials, Imperial College London, London SW7 2AZ, UK}  
\affiliation{Department of Materials Science and Engineering, Yonsei University, Seoul 03722, Korea}

\date{\today}

\begin{abstract}
Halide perovskite semiconductors and solar cells respond to electric fields in a way that varies across time and length scales. 
We discuss the microscopic processes that give rise to the macroscopic polarization of these materials, ranging from the optical and vibrational response to the transport of ions and electrons. 
The strong frequency dependence of the dielectric permittivity can be understood by separating the static dielectric constant into its constituents, including the orientional polarization due to rotating dipoles, which connects theory with experimental observations. 
The controversial issue of ferroelectricity is addressed, where we highlight recent progress in materials and domain characterization, but emphasize the challenge associated with isolating spontaneous lattice polarization from other processes such as charged defect formation and transport. 
We conclude that CH$_3$NH$_3$PbI$_3$ exhibits many features characteristic of a ferroelastic electret, where a spontaneous lattice strain is coupled to long-lived metastable polarization states. 
\end{abstract}

%
%

\maketitle 

\section{Introduction}
Semiconducting halide perovskite materials have attracted intense research interest over the past five years due to the potential for inexpensive solution processing, and desirable optoelectronic properties for photovoltaic (PV) and light emission applications.\cite{Stranks2014,Kojima2009,Gong2018} 
%
%
Halide perovskites synthesize with the chemical formula \ce{ABX3}, where A is a positively charged cation located in the central cavity created by an extended framework of corner-sharing \ce{BX6} metal-halide octahedra.
Thin-film devices based upon the hybrid organic-inorganic perovskite ${\ce{(CH3NH3)}}^+{\ce{PbI3}}^-$ (\ce{MAPbI3}) were the first to utilize a halide perovskite as the PV absorber layer.\cite{Kojima2009,Lee2012}
This material features prominently in the literature due to the rapid increase in power conversion efficiency (PCE), from 3.8$\%$ to 23.3$\%$.\cite{Shin2017,GoogSchol}
%
%
%
However, \ce{MAPbI3} has been shown to be chemically unstable and contains the toxic element Pb.\cite{Yang2016,Lujan2015} 
Therefore, materials with elemental compositions including: 
A = \ce{CH3(\ce{NH2})2}\textsuperscript{+} (FA), Cs\textsuperscript{+}, Rb\textsuperscript{+};\cite{Koh2013,Eperon2015,Protesescu2015,Saliba2016} B = Sn\textsuperscript{2+}, and Ge\textsuperscript{2+};\cite{Noel2014,Hao2014,Gu2000} and X = Br\textsuperscript{-}, and Cl\textsuperscript{-},\cite{Edri2013,Kulbak2016,Hui2014,Chen2015} as well as numerous 2-D layered perovskites,\cite{Gong2018,Tsai2016,Byun2016} have been widely studied.
%
%
The best performing PV devices (PCE and lifetime) are based upon the mixed-cation, mixed-halide perovskite (Cs,MA,FA)Pb\ce{(I{,}Br)3} and are currently the most promising route to commercialize the technology.\cite{Saliba2016Cs,NREL2018,OPV2018}
%
%

%
%
\ce{MAPbI3} has been shown to form in three perovskite structures: a high-temperature cubic Pm$\bar{3}$m phase stable above 330K; a tetragonal phase (I4/mm or I4/mcm) between 330K and 160K; and a final low-temperature orthorhombic phase (Pnma) stable below 160K.\cite{Baikie2013} 
%
%
Similar phase behavior is observed for other halide perovskites, with the phase stability and transition temperatures being influenced by factors including the radius ratio of the chemical constituents.\cite{Stoumpos2013,Kieslich2014,Travis2016}
%
%

%
%
A number of models have been proposed to explain the origin of the high performance of halide perovskite solar cells, including:
defect tolerance;\cite{Berry2015}
Rashba splitting;\cite{Amat2014}
large polarons;\cite{Frost2017} and
ferroelectric (FE) polarons.\cite{Miyata2018}
%
%
Polar nanodomains, observed by piezoforce microscopy (PFM),\cite{Kutes2014,Coll2015,Kim2015,Hermes2016,Rohm2017,Strelcove2017} have also been suggested to contribute towards improved charge separation and transport properties.\cite{Frost2014,Rossi2018}
%
%
%
However, following the confirmation of ionic transport, current-voltage ($J$-$V$) curve hysteresis -- typically an indicator of ferroelectricity -- has been attributed to the migration of mobile ions rather than spontaneous lattice polarization.\cite{Edmands2015,Meloni2016}
Photo-enhanced ionic conductivity has also been reported.\cite{YeongKim2018}
%
%
The most recent effective-circuit models suggest that coupling between electronic and ionic charge at interfaces dominates the charge transport and extraction properties of halide perovskite solar devices, though a consensus is yet to be formed.\cite{Moia2018,Weber2018} 
%
%

%
%
The dielectric response function describes the long-range response of a material to an external electric field.
This can provide information on the underlying physical mechanisms which manifest such diverse behavior.
In halide perovskites, differences in the microscopic crystalline properties of samples, experimental or computational methodologies, and nomenclature lead to disparity in dielectric literature which can be confusing to navigate.
In this Perspective, we briefly introduce the theory of dielectric polarization in order to critically interpret the polarization mechanisms reported for halide perovskites. 
Where possible, we highlight publications which employ the 'best practice' experimental techniques for investigating the dielectric response.
Additionally, we contemplate the ferroic nature of halide perovskites that continues to inspire debate in the field.
\begin{center}
\begin{figure}
	\includegraphics[width=\linewidth]{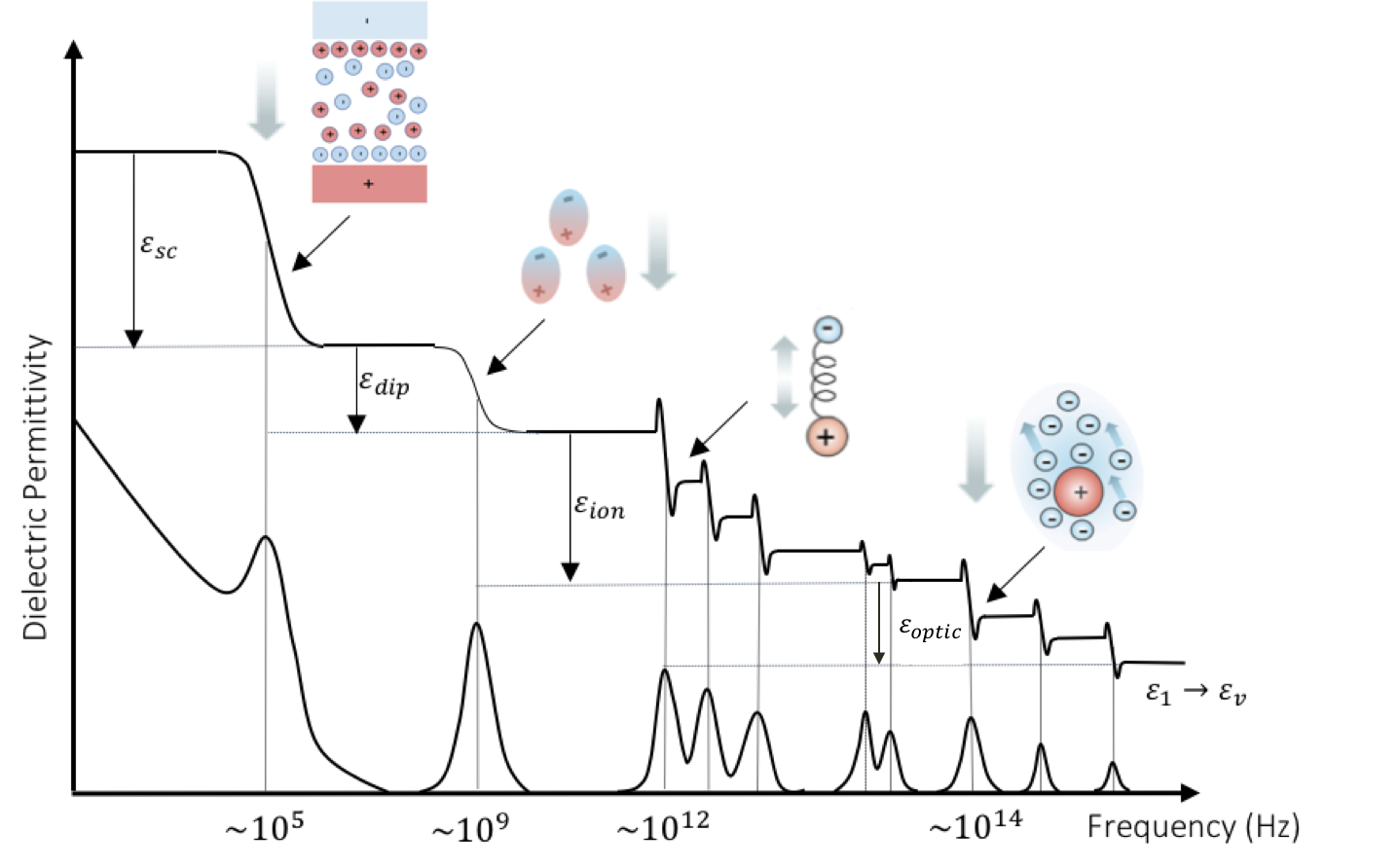}
    \caption{Illustration of the frequency dependent dielectric spectrum typical of halide perovskites. 
	The $x$-axis is scaled to provide an indication of the frequency at which the physical resonances are observed.
    The smooth (top) and piecewise (bottom) curves represent the real ($\epsilon_1$) and imaginary ($\epsilon_2$) components, respectively.
    The inset images schematically demonstrate the physical mechanisms which induce the various components of the static dielectric response ($\epsilon_r$) are described in Equation 1. $\epsilon_v$ refers to the vacuum permittivity. 
    }
    \label{fig:Spectra}
\end{figure}
\end{center}
\section{Dielectric Polarization in Crystals}
A dielectric is defined as a medium which cannot completely screen a static, external, macroscopic electric field from its interior due to physical constraints on charge rearrangement.\cite{Britannica}
The response of a dielectric to an applied electric field is described by the theory of dielectric polarization and is usually framed within classical electrodynamics.\cite{Jackson2012,Zangwill2013}
An external electric field will act to distort the ground-state charge density of a dielectric material with which it interacts –-- schematically shown in the inset diagrams of Figure \ref{fig:Spectra}.
%
%
If the displacement is such that the centre of positive charge no longer corresponds to the centre of negative charge, a polarization field, \textit{P}, is induced; that is, the material becomes polarized.
The polarization field opposes the direction of the external field, the effect of which is to reduce (screen) the electric field present in the bulk material, \textit{E}, when compared to the electric field present in the absence of a dielectric ($E_0$). 
Considering only the dielectric response of the electronic charge density, and in the small, static, linear limit, a dielectric obeys the constitutive equation:\cite{Wooten1972}
\begin{equation} \label{eq:1}
E = \frac{D}{\epsilon(\omega)} = \frac{D - 4\pi{P}}{\epsilon_{v}},
\end{equation}
where, $\epsilon(\omega)$ is the dielectric permittivity, $\epsilon_{v}$ (sometimes written $\epsilon_{0}$), is the vacuum permittivity or, equally, the permittivity of free space, and we have introduced the auxillary vector field, $D = \epsilon_vE_0$.  
The dielectric permittivity is a complex function ($\epsilon(\omega) = \epsilon_1(\omega) + i\epsilon_2(\omega)$) of the frequency of the applied field ($\omega$) that quantifies the linear dielectric response of a material to a constant applied field.
In complex materials, such as the halide perovskites, multiple dielectric mechanisms (physical processes which manifest dielectric polarization) are present.\cite{Zheludev1971}
If the same assumptions as above are taken, the contributions from these mechanisms toward the dielectric permittivity are additive, such that we may write:
\begin{equation} \label{eq:2}
\epsilon_r = \epsilon_{optic} + \epsilon_{ion} + \epsilon_{dip} + \epsilon_{sc},
\end{equation}
where $\epsilon_r = \epsilon/\epsilon_v$ is the relative permittivity and we have dropped the frequency dependency for ease of notation. 
The optical dielectric response, $\epsilon_{optic}$, is due to the (femtosecond) response of the electron density. 
The ionic contribution, $\epsilon_{ion}$, is due to the (picosecond) response of lattice vibrations (phonon modes), and is proportional to the polarity of the chemical bonds and the softness of the vibrations.\cite{Zangwill2013}
%
%
The orientational component, $\epsilon_{dip}$, is due to the slower (nanosecond) realignment of any dipolar species.\cite{Kirkwood1939} 
The space charge contribution, $\epsilon_{sc}$, results from free charges (both ionic and electronic) redistributing (in microseconds to seconds) over macroscopic distances in the material.\cite{Kao2004_Diel}
The theoretical and practical considerations necessary to perform computational or experimental investigations of these processes shall be outlined in the following subsections, though we point the reader to Ref. \onlinecite{Martin2004,Spaldin2012} for in-depth reviews on the core topics. 
%
%
%
%
\subsection{Dielectric polarization in theory}
The dielectric polarization of a solid is often defined as the sum of the induced dipole of a polarizable unit, divided by the volume of the unit.\cite{Ashcroft1976}
%
%
This approach is well defined for finite ionic systems and forms the theoretical foundation of the Clausius-Mossotti model.\cite{Felderhof1983}
%
%
However, it breaks down in the thermodynamic limit (e.g. for extended crystals with delocalized, periodic electronic wavefunctions), as the charge density cannot be unambiguously decomposed into local contributions.\cite{Resta2007,walsh2018oxidation}
%
%

%
%
The modern theory of polarization utilizes the framework of density functional theory (DFT) to contemplate the response of the electron wavefunctions in terms of a geometric (Berry) phase, and not as a charge density.\cite{Kohn1965,Kohn1996}
Consequently, the mathematical formalism which we introduce in the following subsection only addresses the dielectric response of electrons and ions.
The theoretical formalism describing the orientational and the space-charge dielectric response shall be presented in Section \ref{sec:3}.
The quantity of interest in the modern theory is the \textit{change} in polarization, $\Delta{P}$.
It is common in the literature to assume the Born-Oppenheimer approximation in order to separate the change in polarization associated with the response of electrons and ions:\cite{Resta1994} 
%
%
\begin{equation} \label{eq:3}
\Delta{P} = \Delta{P}_{el} + \Delta{P}_{ion}.
\end{equation}
Using DFT one can represent the electronic charge density in terms of the eigenfunctions of the Kohn-Sham (KS) Hamiltonian (the KS orbitals), $\phi_n^{\lambda}$.
The change in electronic polarization, $\Delta{P_{el}}$, that is induced upon an adiabatic transformation  ($\phi_{k,n}^{(\lambda)} \rightarrow \phi_{k,m}^{(\lambda)}$) may then be written:\cite{Resta1992} 
\begin{equation} \label{eq:4}
\Delta{P_{el}} = \int_{0}^{1}d\lambda{P_{el}^{'}(\lambda)},
\end{equation}
where the partial derivative, $P'_{el}(\lambda)$, is taken with respect to $\lambda$ -- a variable of the KS Hamiltonian which parameterizes the transformation, and which is chosen to take the value of zero and one at the initial and final state, respectively. 
If $\lambda$ is taken to be time, for example, then the change in polarization may be cogitated as the integrated polarization current.
The derivative is often stated in terms of first-order density functional perturbation theory (DFPT):\cite{Smith1993} 
\begin{equation} \label{eq:5}
P_{el}^{'}(\lambda) = \frac{-iq_e}{m_e\Omega}\sum_{n}f_{n}\sum_{m\neq{n}}[\frac{\bra{\phi_n}p\ket{\phi_m}\bra{\phi_m}V_{KS}^{'(\lambda)}\ket{\phi_n}}{(\varepsilon_n - \varepsilon_m)^2} + c.c.],
\end{equation}
where $m_e$ and $q_e$ are the electron mass and bare charge, respectively, $\varepsilon_n$ is the Kohn-Sham eigenvalue and $f_n$ is the occupation number of state \textit{n}, $V_{KS}^{'}(\lambda)$ is the derivative of the Kohn-Sham potential with respect to $\lambda$, and \textit{c.c.} refers to charge conjugate terms.
The modern theory assumes that the transition from state $n \rightarrow m$ occurs at null electric field, such that periodic boundary conditions are valid for any value of $\lambda$.
In this regime the KS orbitals take the Bloch form, $\phi_{k,n}^{(\lambda)} \rightarrow u_{k,n}^{(\lambda)} = e^{iGr}u_{k+G,n}^{(\lambda)}$, and Eq. \ref{eq:5} may be recast into a form in which conduction states do not explicitly appear:\cite{Smith1993,Thoulles1982}
%
%
\begin{equation} \label{eq:6}
P_{el}^{'}(\lambda) = \frac{ifq_e}{8\pi^3}\sum_{n=1}^M\int_{BZ}dk\bra{u_{k,n}^{(\lambda)}}\partial/\partial{k}\ket{u_{k,n}^{(\lambda)}}
\end{equation}
The right hand side of Eq. \ref{eq:6} is closely related to the Berry phase of band \textit{n}.\cite{Berry1984,Zak1989}
The modern interpretation of electronic polarization in solids therefore states that $\Delta{P}$ is proportional to the change in Berry phase.
%
%
One can further develop this expression by Fourier transforming the Bloch states to define one-electron Wannier centres, the displacement of which is proportional to the electronic polarization.\cite{Smith1993,Wannier1980}
One may define the electronic dielectric response as the derivative of the polarization with respect to the macroscopic electric field with the nuclei at fixed positions.\cite{Baroni2001}
\begin{equation} \label{eq:7}
\epsilon_{optic}^{ab} = \delta_{ab} + 4\pi\frac{\partial P_{a}}{\partial E_{b}}.
\end{equation}
Here \textit{a} and \textit{b} are lattice directions and $\delta_{ab}$ is the Kronecker-delta.
We have introduced the tensorial form of the dielectric response, $\epsilon_{optic}^{ab}$, as although in the preceding discussion we contemplated the high-symmetry (isotropic) case, a second-rank tensor is required for a general description of all crystal symmetries. 
By removing the restriction of static nuclei, one can calculate the dielectric response including lattice vibrations.
To do so, it is necessary to introduce the Born-effective charge (BEC) tensor:\cite{Butcher2013}
%
%
\begin{equation} \label{eq:8}
Z_{s,ab}^{*}= \frac{\Omega}{|q_e|}\frac{\partial P_a}{\partial r_b^{(s)}},
\end{equation}
which is defined as the linear proportionality coefficient relating the polarization induced in lattice direction, \textit{a}, to a displacement, \textit{r}, of the sublattice, \textit{s}, in the lattice direction, \textit{b}.
The asterisk is introduced to state explicitly that it is an effective charge. 
When combined with knowledge of the phonon eigenmodes, $\omega_0$ (which one can calculate with DFT using finite-displacements or within perturbation theory), the ionic (vibrational) contribution to the dielectric response may be calculated, using:\cite{Gonze1997,Petousis2016}
%
%
\begin{equation} \label{eq:9}
\epsilon_{ion}^{ab} =  \frac{4\pi}{\Omega}\sum_{i=1}^N\frac{(\sum_{sa'}Z^*_{s,aa'}U_i^*)(\sum_{s'b'}Z^*_{s',bb'}U_i)}{\omega_{0i}^2 - \omega^2}.
\end{equation}
Here, the summation is over \textit{N} phonon modes, and $U_i$  are the eigendisplacements of the interatomic force constant matrix, $\tilde{C} = \delta^2E/(\delta{r_{sa}}\delta{r_{s'b}})$, (asterisk denotes the complex conjugate).
Finally, the change in polarization manifest by lattice displacements can be calculated by:\cite{Resta1994}
\begin{equation} \label{eq:10}
\Delta{P}_{ion} = \frac{q_e}{\Omega}\sum_{s}Z^{*}_{s}r_{s},
\end{equation}
%
%
which takes a similar form to Eq. \ref{eq:4}.
The modern theory is commonly used in first-principles calculations of crystalline solids.\cite{Baroni1986,Baroni2001,Resta2018} 
%
%
In general, good agreement is found between the magnitude of spontaneous polarization calculated from Eq. \ref{eq:3} and measurements of conventional ferroelectrics. 
For example, the calculated value of $\Delta P = 0.28$ C/m$^2$ for tetragonal \ce{BaTiO3} compares well with a measurement of $\Delta P = 0.27$ C/m$^2$.\cite{Zhong1994}
%
%
%
%
%
%
%
\subsection{Dielectric polarization in practice}
The dielectric polarization of a crystal cannot be reliably measured as an intrinsic, bulk property.
Generally, one experimentally measures the change in polarization between two polarization states from hysteresis loops that are generated upon switching the direction of polarization currents.\cite{Fukunaga2008}
The mathematical formalism introduced above was motivated, in part, by a desire to compare theoretical calculations with such measurements. 
The majority of studies investigating polarization phenomena today examine its derivatives, however, such as pyro-/piezo-electric coefficients.
In the context of solar cells, measurements of the relative permittivity are frequently reported due to its significance in calculations of physical constants (e.g. absorption coefficient) that impact device performance.
The complete device-relevant dielectric response occurs over a large frequency range (Hz--PHz), requiring multiple complementary experimental techniques (on inevitably different devices) to fully characterize.
Ellipsometry is an optical technique employed to investigate the electronic dielectric response of crystalline samples.\cite{Leguy2016,Peter2010}
%
%
This technique measures the amplitude and phase of (monochromatic) polarized light after reflection off a dielectric surface from oblique incidence.
%
%
Electromagnetic radiation which is transmitted or reflected from oblique incidence has two possible polarization vectors: (1) in which the plane of oscillation is the same for incident and reflected radiation, which we label p-polarization; and (2) in which they are not, which we label s-polarization.\cite{SPpolar}
Upon reflection, the p- and s-polarizations will manifest different changes in amplitude and phase due to the difference in electric dipole radiation.
It is therefore the amplitude ratio ($\Phi$), and phase difference ($\Delta$) between p- and s-polarizations which ellipsometry can provide access to. 
These quantities may be used to determine the amplitude reflection coefficients, $r_p$ and $r_s$, via the relation:\cite{Fujiwara2007}
\begin{equation}
    \rho \equiv \tan{\Phi}e^{i\Delta} \equiv \frac{r_p}{r_s},
\end{equation}
where $\rho$ is the reflectivity.
The complex refractive index, ($\tilde{n} = n + ik$), can subsequently be calculated using the general Fresnel equations, seen below, and mapped to the real and imaginary components of the optical dielectric response via the Maxwell relation, $\epsilon_{optic} = {n}^2$.\cite{agranovich1984, Fox2001A} 
\begin{equation}
\frac{r_p}{r_s} = \bigg(\frac{\tilde{n_1}\cos{\theta_0} - \tilde{n_0}\cos{\theta_1}}{\tilde{n_1}\cos{\theta_0} + \tilde{n_0}\cos{\theta_1}}\bigg)\bigg/\bigg(\frac{\tilde{n_0}\cos{\theta_0} - \tilde{n_1}\cos{\theta_1}}{\tilde{n_0}\cos{\theta_0} + \tilde{n_1}\cos{\theta_1}}\bigg),
\end{equation}
here $\theta_{0}$ and $\theta_{1}$ correspond to the angle of incidence and reflectance, respectively, and $\tilde{n_0}$ and $\tilde{n_1}$ are the refractive index of the incident media and of the dielectric, respectively. 
In conventional semiconductors and insulators, good agreement is found between calculation and measurement of optical dielectric constants. 
For example, DFPT calculations of $\epsilon_{optic} = 13.6$ for Si correspond well with a measurement of $\epsilon_{optic} = 13.8$ from ellipsometry.\cite{Green1995,Jellison1983} 
%
%

%
The dielectric contribution from the response of ions is frequently investigated using infrared (IR) spectroscopy.\cite{stuart2005}
The proportion of light which is transmitted through a dielectric is measured in order to determine the angular frequency of the materials phonon modes. 
The theory of lattice dynamics describes how, at finite temperatures, ions vibrate around their equilibrium crystallographic positions.\cite{maradudin1963}
This motion may be enhanced by photoabsorption; a phenomena which is magnified when the frequency of the perturbing field approaches the natural frequency of the vibrational mode.
The angular frequency of vibrational modes are thus identified as troughs in transmittance spectra.
The oscillation of neighbouring atoms of opposite charge may be in-phase (acoustic phonons) or opposite-in-phase (optical phonons).
Therefore, optical phonon modes induce polarization fields which contribute toward the macroscopic dielectric response, whereas acoustic phonon modes do not.
If the optical dielectric response is known, the angular frequency of optical phonon modes may be related to the ionic dielectric response via the Lyddane-Sachs-Teller relation:\cite{Lyddane1941}
\begin{equation} \label{eq:14}
\frac{\epsilon_{ion}+\epsilon_{optic}}{\epsilon_{optic}} = \frac{\omega_{LO}}{\omega_{TO}}.
\end{equation}
%
%
Here the subscripts LO and TO refer to longitudinal optical (oscillation in the direction of motion) and transverse optical (oscillation perpendicular to the direction of motion) phonon modes, respectively.
In order to determine both LO and TO phonon modes, it is necessary to perform measurements at both normal and oblique incidence.\cite{Berreman1963}
The harmonic approximation is taken in deriving the LST equation. 
Consequently, Eq. \ref{eq:14} cannot accurately describe systems with strong anharmonicity, molecular reorientation, or charge transport.\cite{CHAVES1973}

Materials which feature lower-frequency responses are often labelled `lossy dielectrics', as the associated dielectric mechanisms are not harmonic (resonant) processes, but instead totally dispersive.
As such, optical measurements are not an appropriate probe of this behavior. 
A common interpretation of the relative permittivity is as: 
the ratio of the capacitance of a capacitor whose electrodes are separated by vacuum to a capacitor whose electrodes are separated by a dielectric. 
Impedance spectroscopy, a technique where the dielectric medium is treated as a capacitor, is therefore often employed.\cite{KREMER2002,barsoukov2018impedance}
%
%
This technique requires many considerations and involves complex analysis; the interface between the dielectric and the electrode contacts being an important factor.\cite{GERHARDT1994,COSTER1996}
By performing the techniques introduced above over a range of frequencies (spectroscopic measurements) one can produce a dielectric spectrum as schematically shown in Fig. \ref{fig:Spectra}.
Discounting resonance effects, the complex refractive index exhibits normal dispersion (monotonically increases as the applied frequency decreases).
Resonant behavior dominates when the frequency of the applied field approaches the natural frequency of the underlying dielectric process.\cite{Blythe2008} 
This gives rise to a characteristic peak in the imaginary component (a Lorentzian), and a step function in the real component, as seen in Figure \ref{fig:Spectra}.  
If the frequency of the applied field is above the response time of the process, it cannot respond fast enough to contribute to the screening. 
Once a spectrum has been measured, contributions from individual dielectric processes can be assessed within the Debye relaxation model:\cite{Debye1929}
%
%
\begin{equation} \label{eq:13}
\tilde{\epsilon}(\omega) = \epsilon_{optic} + \frac{\epsilon_s - \epsilon_{optic}}{1 + i\omega\tau_d}.
\end{equation}
Here $\tau_d = \omega^{-1}_0$ is the dielectric relaxation time, and $\epsilon_s = \epsilon_{optic} + \epsilon_{ion} + \epsilon_{dip}$, is the static dielectric response.
Extensions of the model have been proposed in order to account for the finite width of the relaxation time distribution, and non-linearity in the high-frequency regime present in devices.\cite{Cole1941,Cole1942,davidson1950,HAVRILIAK1967}
Responses can overlap in frequency (e.g. near optical transitions) or may have such broad frequency response that they exceed the measurement window.
This makes unambiguous interpretation difficult.
Consequently, few materials have a complete dielectric function characterization; fullerenes are one case where a full spectrum is known.\cite{Eklund1995}
\section{Dielectric properties of Halide Perovskites} \label{sec:3}
\subsection{Optical dielectric response}
Optical absorption involves the photoexcitation of electrons from valence to conduction bands; creating either bound electron-hole pairs (excitons) or free carriers. 
As such, the optical dielectric response is dominated by the optical band gap and other low energy band-edge states. 
For MAPb\ce{I3}, it has been deduced that the valence and conduction bands are composed of hybridized I 5p orbitals and Pb 6p orbitals, respectively.\cite{Butler2015}
%
%
The optical band gap corresponds to the VB1 $\rightarrow$ CB1 transition at the $R$ symmetry point, and has been measured at 1.6eV ($\sim 400$ THz).
Further excitonic absorption has been suggested at 2.48 eV (600 THz) and 3.08 eV (745 THz).\cite{Leguy2016}
At room temperature, the potential energy surface for the orientation of the \ce{CH3NH3+} ion within the lead-iodide octahedra is soft.
Consequently, there is computational sensitivity to the choice of electronic structure Hamiltonian, and level of geometry optimization.\cite{Zhou2015,Brivio2014} 
%
%
However, DFT calculations applying Eq. \ref{eq:7} within the generalized gradient approximation produce an isotropically averaged value of $\epsilon_{optic} = 6.0$ for a crystal with MA aligned in the low energy $\braket{100}$ direction.\cite{Brivio2013} 
%
%
This result was replicated by Ref. \onlinecite{Zhou2015}, whom utilized similar methods.
These calculations compare well with the `best practice' experimental value, $\epsilon_{optic} = 5.5$; produced from ellipsometry measurements on single crystals.\cite{Leguy2016}
Leguy et al. also measure the optical response of \ce{MAPbI3} thin films, and suggest that the reduction in $\epsilon_{optic}$, from $5.5 \rightarrow 4.0$, is due to surface effects.\cite{Leguy2016}
A larger response (6.5) is reported by Hirasawa et al., though we suggest this is due to assumptions taken in data processing and not to an increase in the dielectric response.\cite{Hirawasa1994}
Further, we posit that Ref. [\onlinecite{Glaser2015}] underestimate the electronic response ($5.0 \leq 5.5$) due to the classical Lorentz dipole fitting model which is employed.
To support this statement, when additional Debye components are included in a later study which employed similar methods, a value closer to the consensus is calculated (5.5).\cite{Chavez2015}
%
%

%
%
The polarizability of a compound, and hence the electronic dielectric response, is inversely proportional to the magnitude of its band gap, as described by second-order perturbation theory.\cite{PerturbTheory}
%
%
For halide perovskites, a number of general trends related to changes in the band gap can therefore be observed in the published values of $\epsilon_{optic}$ summarized in Table \ref{table:1} for different \ce{ABX3} compounds.
The decrease in reported values of $\epsilon_{optic}$ on transition from I (5p) to Br (4p) to Cl (3p) has been understood by considering the higher binding energy of valence electrons and the corresponding increases in optical band gap ($E_g^{I} \sim 1.6$ eV $\rightarrow E_g^{Br} \sim 2.2$ eV $\rightarrow E_g^{Cl} \sim 2.9$ eV).\cite{Shi2015,Maculan2015,Bokdam2016}
%
%
Spin-orbit coupling (SOC) plays an important role in determining the conduction band for compounds which include Pb ($Z$ = 82) in their composition.
Exchanging Pb for Sn ($Z$ = 50) leads to a decrease in the optical band gap ($E_g^{Pb} \sim 1.6$ eV $\rightarrow E_g^{Sn} \sim 1.2$ eV) and to the expected increase in $\epsilon_{optic}$.\cite{Umari2014,Stoumpos2013,Whalley2017}
%
%
While the A-site cations do not directly contribute to the band edge states, they do influence the crystal structure and metal-halide bond strength.\cite{Amat2014}
%
%
The volumetric decrease on exchanging MA for \ce{Cs} leads to an increase in the optical band gap ($E_g^{MA} \sim 1.6$ eV $\rightarrow E_g^{Cs} \sim 1.7$ eV), and to a decrease in $\epsilon_{optic}$, again.\cite{Eperon2015,Stoumpos2013,Leguy2016}
%
%
%
%
\begin{table}
\caption{Representative values from computational (C) and experimental (E) studies reported in the literature for the individual contributions to the total dielectric response from the polarization mechanisms present in halide perovskites. For publications in which the individual contributions were not given, the values from the `best practice' method (underlined) associated with higher-frequency processes has been deducted. An asterisk denotes a value cannot be separated into individual contributions.}
\small
\centering
\begin{tabular}{ p{1.5cm} p{0.8cm} p{0.8cm} p{0.8cm} p{0.8cm}  p{0.8cm} p{0.8cm} p{1.4cm}}
\hline\hline
\multirow{2}{*}{Material}  & \multicolumn{2}{p{1.5cm}}{  $\epsilon_{optic}$} & \multicolumn{2}{p{1.5cm}}{  $\epsilon_{ion}$} & \multicolumn{2}{p{1.5cm}}{  $\epsilon_{dip}$} & 
 {  $\epsilon_{sc}$} \\ [1ex]
 & C & E & C & E & C & E & E \\ 
\hline 
MAPb\ce{I3} 
& 4.5\cite{Brivio2014} 5.1\cite{JIANG2015,Du2014} 5.3\cite{Proupin2014} 5.8\cite{Osorio2015} 6.0\cite{Brivio2013,Zhou2015,MA2015} 6.8\cite{Bokdam2016} 7.1\cite{Umari2014} 
&  4.0\cite{loper2014complex} 5.0\cite{Glaser2015} \underline{5.5}\cite{Chavez2015,Leguy2016} 6.5\cite{Hirawasa1994}
& 16.6\cite{Du2014} 16.7\cite{Brivio2013} 23.2\cite{Bokdam2016} 
& \underline{16.5}\cite{Govinda2017} 17.8\cite{YAMAMURO1992} 23.3\cite{Poglitsch1987} 24.5\cite{Anusca2017} 28.5\cite{Sendner2016}
& 8.9\cite{Frost2017}
& 13\cite{Lin2015} 32.1\cite{Green2014} 32\cite{Anusca2017} 36.9\cite{Green2014} 
& 35\cite{Lin2015}    $10^{3-7}$\cite{Perez2014}
\\
MAPb\ce{Br3}
& 5.2\cite{Bokdam2016} 6.7\cite{Zhao2017} 
& \underline{4.0}\cite{Leguy2016} 4.7\cite{Glaser2015} 4.8\cite{TANAKA2003} 
& 18.3\cite{Zhao2017}
& \underline{16.0}\cite{Govinda2017}  21.5\cite{Poglitsch1987} 24.7\cite{YAMAMURO1992} 27.6\cite{Sendner2016}
& 
& 38\cite{Anusca2017}

\\ 
MAPb\ce{Cl3}
& 4.2\cite{Bokdam2016}
& \underline{3.1}\cite{Leguy2016} 4.0\cite{Glaser2015}  
& 
& 11.9\cite{Anusca2017} \underline{19.0}\cite{YAMAMURO1992} 20.8\cite{Poglitsch1987}  25.8\cite{Sendner2016}
&
& 30\cite{Anusca2017}
\\ 
MASn\ce{I3} 
& 8.2\cite{Umari2014}
\\ 
CsPb\ce{I3} 
& 5.3\cite{Brgoch2014}
\\  
CsPb\ce{Br3} 
& & &
& 20.5*\cite{Govinda2017}
\\
\hline\hline
 \label{table:1} 
\end{tabular}
\end{table}
\subsection{Ionic dielectric response}
A description of the position of ions within the lattice at finite temperature involves an interplay between thermal motion and interatomic restoring forces (e.g. hydrogen bonding and Van der Waals forces).
It is common to assume that the motion is elastic and non-dissipative -- defined as the harmonic approximation.
Anharmonic interactions can also occur, however, and have been suggested in \ce{MAPbI3} due to Pb off-centering and octahedral tilting.\cite{Whalley2016,Young2015}
%
%
%
Despite this, the harmonic phonon dispersion has been fully characterized for \ce{MAPbI3}; and is found to be dominated by vibrations of the \ce{PbI6} octahedra from $0.5-3.2$ THz.\cite{Osorio2015,leguy2016dynamic}
The low energy of these modes can explain why values for $\epsilon_{ion}$ seen in Table \ref{table:1} for the halide perovskites are much larger than for tetrahedral semiconductors such as \ce{CdTe} ($\epsilon_{ion} = 2.5$).\cite{Strauch2012}
%
%
The strength of the ionic polarization has been suggested to enhance PV performance (e.g. through defect tolerance) and should be an important consideration in future material searches.\cite{Wehrenfennig2013,Brandt2015,Ganose2017,walsh2017instilling}
The averaged BEC tensors for \ce{MAPbI3} have been calculated to be larger than the formal ionic charges ($Z_{Pb}^* = +4.42$, $Z_{I}^* = -1.88$ $Z_{MA}^* = +1.22$), such that small ion displacements result in large changes to the polarization.\cite{Osorio2015,Du2014}
%
%
Anisotropy in the calculated BEC tensors indicates a preferential direction of vibration for apical and equatorial iodine ions, which has been interpreted as octahedral `breathing'.\cite{Du2014}
%
%
Brivio et al. report an isotropically averaged value of $\epsilon_{ion} = 16.7$ for the case when MA is oriented in the low energy $\braket{100}$ direction, which rises to $\epsilon_{ion} = 26.4$, when MA is oriented in the $\braket{111}$ direction.\cite{Brivio2013}.
%
%
Whilst it is tempting to average over all possible orientations, as done in Ref. \onlinecite{Bokdam2016}, we report values associated with MA aligned in the $\braket{100}$ direction, when possible. 
These calculations compare well with the `best practice' experimental value of $\epsilon_{ion} = 16.5$, determined from impedance measurements performed on powder samples.\cite{YAMAMURO1992}
%
%
In order to account for the dielectric contribution from interfacial polarization (to be introduced in Section III.D), the authors introduce additional Maxwell-Wagner terms to the Debye relaxation model with which they perform their spectral fitting.\cite{Govinda2017,Sillars1937} 
%
%
The neglect of these effects can explain the larger value for $\epsilon_{ion}$ (23.0) obtained by Poglitsch et al. in an earlier study.\cite{Poglitsch1987}
%
%
Sendner et al. report a value of $\epsilon_{ion} = 28.5$ (which rises to 31 if $\epsilon_{optic} = 5.5$ is used) after identifying TO (0.96 THz and 1.9 THz) and LO (1.2 THz and 4.0 THz) phonon modes, and applying the Cochran-Cowley expression.\cite{Sendner2016}
The Cochran-Cowley expression is a generalization of Eq. \ref{eq:14} that accounts for systems with more than two atoms in the unit cell (\ce{MAPbI3} has 48 at room temperature).\cite{COCHRAN1962}
Whilst this value may improve if a greater number of optical phonon modes are included in the calculation (\ce{MAPbI3} has 141 at room temperature), it exemplifies the limitations of the model for describing complex materials.
%
%
%
%

\begin{figure}
	\includegraphics[width=0.75\linewidth]{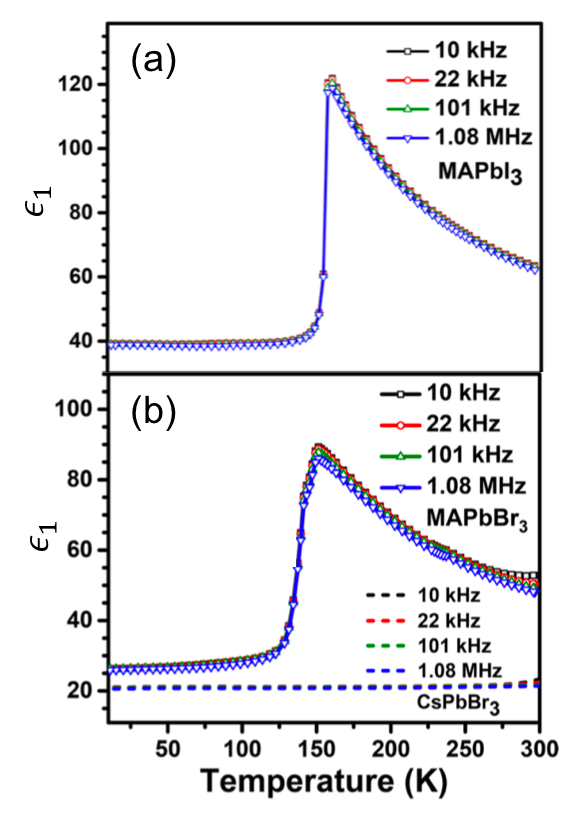}
    \caption{Real dielectric permittivity, $\epsilon_1$, versus temperature for selected frequencies measured by impedance spectroscopy on (a) \ce{MAPbI3} and (b) \ce{MAPbBr3} (symbol and solid lines) and \ce{CsPbBr3} (dashed lines). Adapted with permission from Ref. \onlinecite{Govinda2017}. Copyright 2017 American Chemical Society.}
    \label{fig:TempDep}
\end{figure}

%
The ionic dielectric response is dependent upon the frequency of vibrational modes and the associated ionic charges.
Therefore, compositions containing lighter elements may be expected to exhibit a weaker response due to faster vibrations and less polarizable ions.  
Contemplating the role of the halide, Ref. [\onlinecite{Poglitsch1987}] measure a systematic decrease of $\epsilon_{ion}$ for the sequence MAPb(I)$_3$ $\rightarrow$ (Br)$_3$ $\rightarrow$ (Cl)$_3$.
The authors attribute this behavior to the blue-shift of vibrations associated with the decrease in mass of the halide; an argument supported by Sendner et al.\cite{Sendner2016}
%
%
Whilst, phonon modes greater than 10 THz associated with \ce{CH3NH3} molecular vibration are influential in phase transitions (as seen in Fig. \ref{fig:TempDep}), Bokdam et al. suggest that they have little impact on the ionic dielectric response.\cite{Bokdam2016}
The measurement of similar responses for both \ce{MAPbI3} and \ce{CsPbI3} by Ref. [\onlinecite{Govinda2017}] supports this claim.
%
%
Pb$^{2+}$ is highly polarisable due to its lone pair (6s$^2$) electrons, and is therefore expected to dominate the ionic dielectric response.\cite{Ganose2017}
Lone pair activity with dynamic structural distortions has also been confirmed in Sn$^{2+}$-based halide perovskites.\cite{fabini2016dynamic}
%
%
%
%
%
%
%
\subsection{Orientational dielectric response}
Orientational polarization emerges from the stimulated reorientation of localized dipoles in the bulk of a dielectric. 
The theoretical description was developed within the context of polar liquids and gases, but the same phase physics can be applied to crystalline materials that contain a dipolar species with rotational degrees of freedom.\cite{Wilson1939} 
%
%
The static permittivity for a polar liquid is described by the Kirkwood-Fr\"{o}hlich equation,\cite{Kirkwood1939, Frohlich1968}
%
%
\begin{equation} \label{eq:15}
\frac{(\epsilon_s - \epsilon_{\infty})(2\epsilon_s + \epsilon_{\infty})}{\epsilon_s{(\epsilon_{\infty} + 2)}^2} = \frac{1}{9}\frac{\mu^2N}{\epsilon_vk_bT'}(1 + z cos(\gamma))
\end{equation}
where \textit{N} is the dipole number density, \textit{z} is the number of nearest-dipolar neighbours, $\gamma$ is the angular separation of neighboring dipoles, $\epsilon_v$ is the vacuum permittivity, and $k_b$ is the Boltzmann constant. 
$T'$ is sometimes defined as an effective temperature ($T' = T - T_C$, where $T_C$ is the Curie temperature) to account for dipole-dipole interactions.\cite{Govinda2017,YAMAMURO1992,Bell1964} 
%
%
Here, $\epsilon_{\infty} = \epsilon_{optic}$ for liquids; however, when applied to crystals it also contains the ionic contribution $\epsilon_{ion}$.
The electrical dipole moment, $\mu$, is the Maclaurin series expansion of the applied field with respect to the internal field (though the first order `unperturbed' static value is commonly used in calculations).\cite{Hurst1988} 
This value can differ from the effective dipole moment that governs dipole-dipole interactions due to screening by the encasing polarizable medium.\cite{Wilson1939} 
%
%

%
%
Unlike the `dynamic' dielectric mechanisms previously introduced, the orientational component is not expected to impact electron transport.\cite{Dinpahjooh2017}
%
%
Whilst the reasoning for this effect is beyond the scope of this perspective, it can be evidenced by the correspondence between reported values of charge mobility measured for organic and inorganic halide perovskites.\cite{Herz2017}
%
%
Consequently, the majority of studies contemplating the role A-site dynamics on PV functionality do so within the context of ferroelectricity (landscape of polar domains), and not dielectrics.
However, the orientational component can influence the slower motion of mobile ions and the screening of point defects.\cite{zwanzig1963dielectric, chen2016extended}
%

%
%
Application of Onsager theory for a polar liquid estimates the response due to the \ce{CH3NH3+} dipoles ($\mu = 2.29$ D) to be $\epsilon_{dip} = 8.9$ at T = 300K.\cite{Frost2017}
%
%
This value corresponds well with a multi-approach measurement on thin films, which reports $\epsilon_{dip} \sim 12$.\cite{Du2014}
We required an effective dipole moment of $\mu \sim 0.7$ D in order to reproduce a similar value using Equation \ref{eq:15}, however.
%
%
Reported values of $\mu = 0.85$ D and $\mu = 0.88$ D, derived by fitting spectral data with Equation \ref{eq:15}, are also smaller than theoretical calculations of the static dipole for \ce{CH3NH3+}.\cite{Poglitsch1987,YAMAMURO1992}
This disparity can arise from interactions with the environment or inter-molecular correlation that are not properly described by the polar liquid model.
In contrast to the values above, Anusca et al. assign a stronger orientational response ($\epsilon_{dip}=32$) from impedance measurements on single crystals.\cite{anusca2017dielectric}
The preceding discussion demonstrates the uncertainty in the field and the need for further investigation and methodological developments. 
%
%
\begin{figure*}{t!}
	\includegraphics[width=0.8\textwidth]{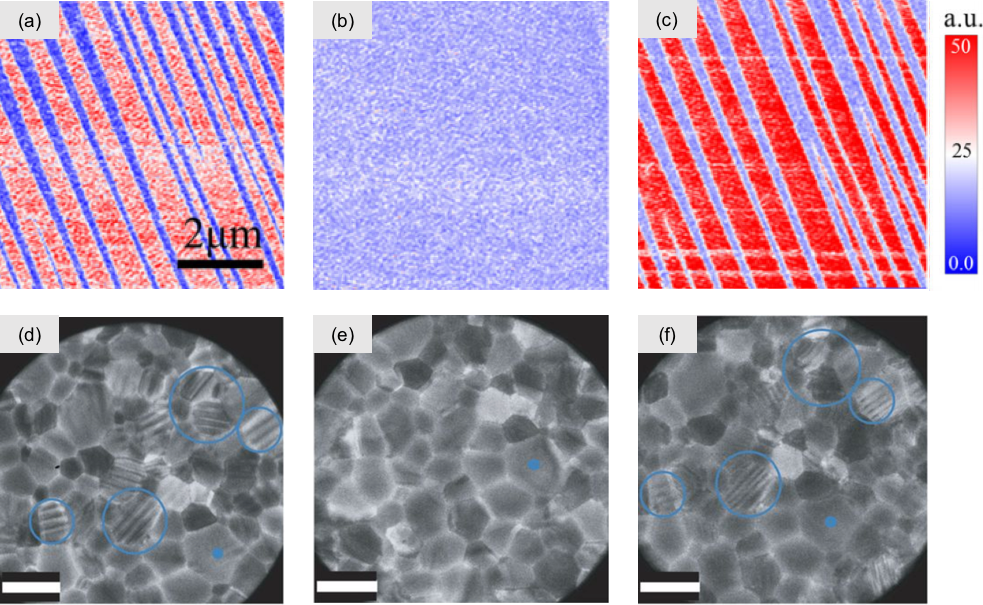}
    \caption{Micrographs showing the evolution of ferroelastic twin domain structures heating (a$\rightarrow$b and d$\rightarrow$e) and cooling (b$\rightarrow$c and e$\rightarrow$f) the cubic-tetragonal phase transition. 
    (a-c) Vertical piezoelectric amplitude measured by piezoforce microscopy (PFM) on a single crystal of \ce{MAPbI3}. Adapted with permission from Ref. \onlinecite{Huang2018}. CC Attribution 4.0 IPL.\cite{CCLicence} 
    (d-f) Bright-field transmission electron microscopy (TEM) on thin film samples.  
    Scale bar is $1\mu$m.  Adapted with permission from Ref. \onlinecite{rothmann2017}. CC Attribution 4.0 IPL.\cite{CCLicence}
    Both samples were annealed before data collection and both PFM and TEM images indicate domains of width $100-300$nm.}
    \label{fig:Ferro1}
\end{figure*}
\subsection{Space charge dielectric response}
Beyond the bulk polarization processes previously discussed, 
understanding of the dielectric response due to the distribution and transport of charged species (molecules, ions, electrons, and holes) is essential to describe and characterize the operational behavior of PV devices. 
As the name suggests, space-charges are formed by \textit{spatial} inhomogeneity in the \textit{charge} distribution which create electrostatic potential gradients. 
The nature of the space charge formed in a device will depend on the processing history and state of a device (e.g. applied voltage).\cite{maier2004physical}
Extended defects -- including surfaces, interfaces, and grain boundaries -- often act as traps for charged species in semiconductors and can support concentrations of charge carriers and defects well above the bulk equilibrium values. 
The distribution of charge can enhance the strength of dielectric polarization through the formation of electrical double layers (EDLs) -- shown in the inset on the far right of Figure 1.
The EDLs behave as conventional capacitors to induce static fields which screen the interfacial charge, and can therefore significantly impact impedance measurements.\cite{Ishai2013}
%
%

%
%
A formal description of space-charge formation requires a solution of the Poisson-Boltzmann equation, given below rearranged in terms of relative permittivity, which is often performed under various approximations (e.g. Helmholtz, Gouy-Chapman,  Stern).\cite{Butt2006,helmholtz1853,Gouy1910,Gouy1917,Chapman1913,stern1924}
\begin{equation} \label{eq:16}
\epsilon_{r} = \frac{c q_e \sinh(\frac{q_e\Phi}{k_BT})}{\epsilon_v\Delta\Phi}.
\end{equation}
Here $\Delta\Phi$ is the Laplacian of the electrostatic potential, and $c$ is the concentration of charge carriers with charge $q_e$.
The space charge is an integral part in drift-diffusion simulations of operating devices.
%
%
The inherent complexity of real devices -- including interfacial inhomogeneity, high defect densities, and phonon scattering -- limits the predictive power of Eq. \ref{eq:16}.
Evidence for space charge polarization instead arrives from observation of Jonscher's law ($\epsilon_2 \propto \omega^{-1}$) in low-frequency dielectric spectra ($\omega < 100$ kHz).\cite{Jonscher1981}
%
%
Jonscher's law can be understood by relating the imaginary dielectric constant to the components of the refractive index, $\epsilon_2 = 2nk$.
The real refractive index, $n = c/v$, describes changes in the phase velocity of electromagnetic propagating through a material.
The extinction coefficient, $k=c\alpha/2\omega$, describes radiation attenuation due to its proportionality to the Beer-Lambert absorption coefficient, $\alpha$.
Observation of Jonscher's law in spectroscopic measurements of \ce{MAPbI3} therefore suggests that lossy processes, such as ion migration, dominate at low frequency.\cite{Du2014,Anusca2017}
Presenting an equivalent circuit model, Moia et al. suggest that the accumulation of charged ionic species at the perovskite-electrical contact interfaces modulates the energetic barrier to charge injection and recombination.\cite{Moia2018}
%
%
It is expected that this effect shall be significantly enhanced under illumination due to an increase in the concentration of free carriers and point defects due to photoexcitation.\cite{YeongKim2018}
As previously mentioned, interfacial charge can impact spectroscopic impedance measurements. 
Consequently, coupling between electronic and ionic transport has been reported to explain the `photoinduced giant dielectric constant' ($\epsilon_r^{dark}\sim 10^3 \rightarrow \epsilon_r^{illu}\sim 10^7$) measured at low frequencies under illumination.\cite{Perez2014,Weber2018}
%
%
%

%
%
In addition to macroscopic space charge effects, hopping polarization emerges from the transition of localized charges ($q$) between electrostatically \textit{inequivalent} lattice sites.\cite{LEAL1981}
%
%
The hopping polarizability, $\alpha_h$, can be calculated by\cite{Kao2004_Diel}
\begin{equation} \label{eq:17}
\alpha_h = \frac{q^2r^2}{3k_BT}\overline{P_{A\rightarrow B}}\overline{P_{0,B\rightarrow A}},
\end{equation}
where $\overline{P_0}=\mathcal{C}exp^{-E_a/k_bT}$ is the ensemble average transition probability at thermal equilibrium, $\mathcal{C}$ is a prefactor containing an attempt frequency, \textit{r} is the spatial separation, and $E_a$ is the transition activation energy.
%
%
When treated as a classical process, the associated dielectric contribution can be approximated via the Clausius-Mossotti relation,
\begin{equation} \label{eq:18}
\frac{\epsilon_r + 1}{\epsilon_r + 2} = \sum_{i=1}^M\frac{N_i\alpha_i}{3\epsilon_v},
\end{equation}
where the sum is over \textit{M} charged species. 
%
%
However, for systems which feature both space-charge and hopping polarization, it is difficult to analytically separate individual contributions.
Consequently, there remains uncertainty regarding the spatial distribution of mobile ions in halide perovskites and the pathways which they migrate through (bulk and surface transport or grain boundaries). 
Similar hopping processes may also occur for electron transport.
Although electrons and holes exist in the form of diffuse large polarons in the bulk materials,\cite{Frost2017} localization and hopping through an inhomogeneous potential energy landscape associated with charged point and extended defects is likely as in the case of H and V-centres.\cite{whalley2017h}
Again, further investigation on this topic is necessary.
\section{Ferroic properties of halide perovskites}
\subsection{Are halide perovskites ferroelectric?}
In analogy to ferromagnets, ferroelectric materials exhibit a \textit{spontaneous} and \textit{reversible} polarization below a characteristic Curie temperature. 
Some of the highest performing ferroelectric materials -- as defined by the magnitude of the spontaneous polarization, the Curie temperature, and the coercive electric field -- are oxide perovskites.\cite{Yang2010}
%
%
In systems such as \ce{BaTiO3} and \ce{PbZr_xTi_{1-x}O3}, spontaneous polarization arises from a displacement of the A or B species from their ideal (cubic) lattice sites.\cite{Cohen1992} 
Credible ferroelectricity has also been reported in certain halide perovskites; below 563K \ce{CsGeCl3} adopts a non-centrosymmetric rhombohedral perovskite structure with lattice polarization along the $\braket{111}$ direction, and exhibits a non-linear optical response.\cite{QINGTIAN2000} 
%
%

%
%
Lossy dielectrics with mobile ions can exhibit apparent ferroelectric signatures. 
This is exemplified by reports of hysteresis cycles produced from electrical measurements on a banana.\cite{Scott2008} 
%
%
We note that stoichiometry gradients (Hebb-Wagner polarization) in mixed ionic-electronic conductors represent a more complicated case.\cite{maier2004physical}
When combined with the molecular dipole, this makes classification of the polar nature of \ce{MAPbI3} difficult.
This uncertainty has led to conflicting reports, such that the tetragonal phase of \ce{MAPbI3} has been referred to as:
`superparaelectric, consisting of randomly oriented linear ferroelectric domains' from Monte Carlo simulations;\cite{Frost2014}
`a ferroelectric relaxor' from dielectric dispersion measurements;\cite{GUO2016}
`ferroelectric' on the basis of electrical measurements on single crystals;\cite{Rakita2017}
and `non-ferroelectric' from analysis of impedance spectroscopy.\cite{Edmands2015}
%
%
Interpreting such disparate results is complicated by differences in experimental technique and sample quality. 
It has been deduced that the \ce{CH3NH3}$^+$ dipoles have a fixed, anti-aligned orientation in the low-temperature orthorhombic phase of \ce{MAPbI3} due to strong hydrogen bonding between the amine group and iodine ions.\cite{Svane2017}
%
%
The crystal structure is centrosymmetric and anti-polar.\cite{Chen2015}
Additional degrees of freedom are introduced upon an increase in the temperature which weaken N\ce{H3}-I bonds and that enable rotational motion of the organic cation.\cite{Weller2015}
%
%
Rotational motion induces structural deformations of the inorganic framework due to an asymmetry in the bond strength between iodine and the two functional groups of the \ce{CH3NH3}$^+$ molecule.
Large polarization currents are induced which dramatically impact the dielectric response, as seen in Figure 2(a), and which drive the tetragonal-to-orthorhombic phase transition at low temperature.\cite{Birkhold2018,Gottesman2014} 
%

%
%
In the tetragonal phase, the \ce{CH3NH3}$^+$ dipole can occupy several fixed orientations within the lattice; the transition between which may be stimulated by thermal motion, for example. 
Govinda et al. argue that the 1/T dependency of the dielectric response, observed above $160$K in Figure 2(a), implies rotational disorder in the \textit{ab} plane.\cite{YAMAMURO1992,Govinda2017}
%
%
The established dynamic disorder of molecular orientation and inorganic octahedral titling evidences the absence of true ferroelectricity in \ce{MAPbI3}.
Following the growing evidence which suggests that polar domains are neither long lived or stable, we suggest that the polar behavior of tetragonal \ce{MAPbI3} exhibits many features common to an electret with a combination of lattice, defect, and surface polarization.
An electret is defined as a dielectric that contains a mixture of surface or bulk charge, which may be due to real charges or dipoles, and under the effect of an applied voltage decays slowly over time.\cite{Kao2004_Electret}
Although one can find reports which claim to contradict this statement, the supporting evidence is consistently inconclusive upon close inspection and comparison between reports.
Beyond hybrid organic-inorganic perovskites that contain polar cations, similar behavior is observed in inorganic halide perovskites (such as CsPb\ce{Br3} and CsPb\ce{I3}) due to \textit{fluctuating} rather than \textit{permanent} electric dipoles.\cite{Kao2004_Diel}
%
%
%
%

%
%
%
%
%
\begin{figure}[t!]
	\includegraphics[width=0.9\linewidth]{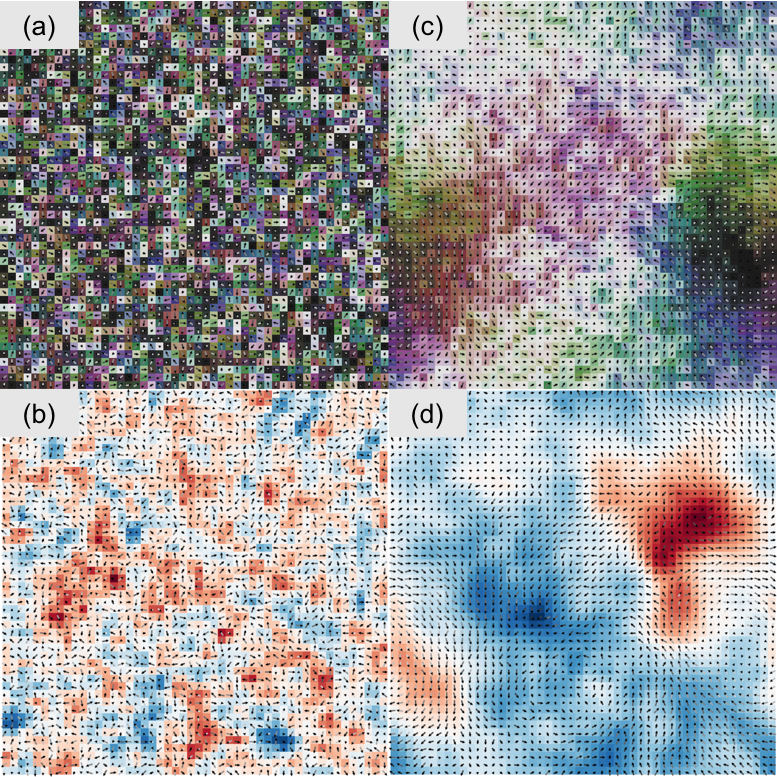}
    \caption{Calculated polarization in \ce{MAPbI3} at room temperature from on-lattice Monte Carlo simulations using the \textsc{Starrynight} code.\cite{StarryNight} Each arrow represents the polarization vector in a perovskite unit cell of magnitude 2.29 Debeye. 
 (a) Dipole orientation and (b) electrostatic potential from a model Hamiltonian containing a screened electrostatic interaction with zero cage strain.
 (c) Dipole orientation and (d) electrostatic potential including a cage strain of 3$k_B$T.
 The colors represent the orientation of the polarization vector (a,c) and the magnitude of the electrostatic potential (b,d) from high (red) to low (blue).
   }
    \label{fig:Starry}
\end{figure}
\subsection{Are halide perovskites ferroelastic?}
A ferroelastic material exhibits a spontaneous and reversible \textit{strain} following a stress-induced phase transition.
The formation of domains occurs in order to minimize the total strain within the material, as established for \ce{BaTiO3}.\cite{Arlt1990,CHOU2000}
%
%
The most convincing evidence for ferroic behavior in \ce{MAPbI3} thin films comes from observation of domain structures in piezoforce microscopy (PFM) amplitude and transmission electron microscopy (TEM) images, shown in Figure 3.\cite{Huang2018,rothmann2017} 
%
%
These have been classified as ferroelastic twin domains.
\begin{figure}
	\includegraphics[width=0.98\linewidth]{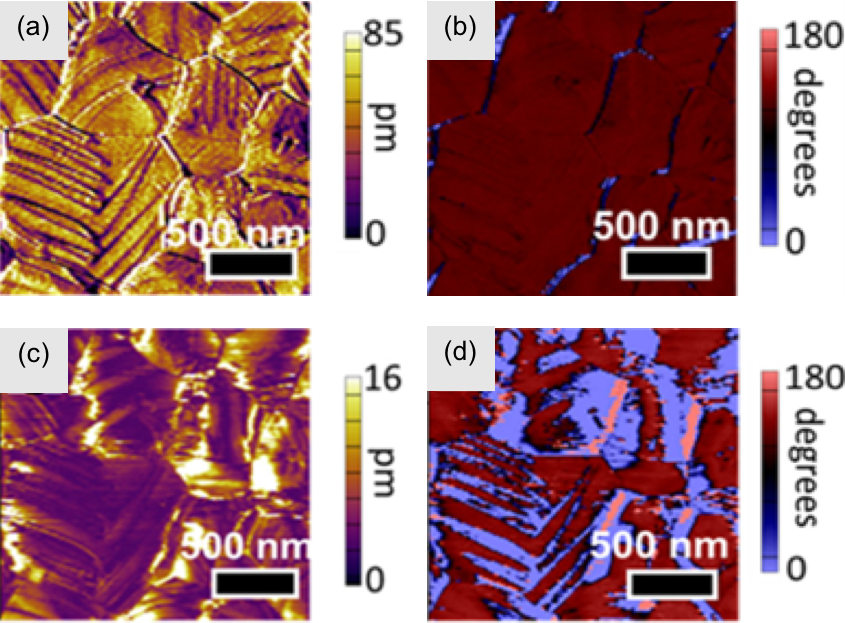}
    \caption{Micrographs from different piezoforce microscopy (PFM) measurements of the same location on a \ce{MAPbI3} thin film sample, including: (a) vertical amplitude (b) vertical phase, (c) lateral amplitude, and (d) lateral phase. The sample was annealed before data collection and displays domains of width $100$nm--$300$nm. Adapted with permission from Ref. \onlinecite{Vorpahl2018}. Copyright 2018 American Chemical Society. }
    \label{fig:Ferro2}
\end{figure}
%
%

%
%
`Ferroelastic fingerprints' were \textit{revealed} by early PFM measurements on these materials.\cite{Hermes2016,Strelcove2017} 
Methodological refinements improved the precision of the PFM technique, such that a transformation of the domain structures upon heating and cooling through the cubic-to-tetragonal phase transition can be conclusively observed in Ref. \onlinecite{Huang2018,Vorpahl2018}.
In analogy to \ce{BaTiO3}, this suggests that the cubic phase is prototypic, and that the cubic-to-tetragonal phase transition is stress induced.\cite{Aizu1969}
Observation of the same structures and behavior in TEM, shown in Figures 3(d-f), implies that the domain structures extend into the bulk material.\cite{rothmann2017}
Moreover, they confirm that the measured PFM signal is not merely due to polarization fields. 
Rothmann et al. suggest that such structures `eluded observation so far [...] due to their very fragile nature under the electron beam' and therefore utilized a low electron dose rate ($\approx 1 e$\AA$^{-2}s^{-1}$) during their TEM experiment to mitigate these effects.\cite{rothmann2017}
%
%
Further support for the assignment of \ce{MAPbI3} as ferroelastic arrives from reports of a $0.5\%$ lattice constriction,\cite{Ren2016} stress induced modifications to domain wall positions,\cite{Strelcove2017}, and periodic differences in the resonant frequency of MA$^+$ dipolar oscillations.\cite{Liu12018,Liu2018,Huang2018}
%
%

%
%
It is clear that whilst there is sufficient evidence to support the assignment of the domains forming in \ce{MAPbI3} as ferroelastic, that these alone cannot explain the full range of observed behavior.  
After all, ferroelastic ``domains cannot be probed by PFM'', as R\"{o}hm et al. points out.\cite{Rohm2017}
However, as seen for the case of \ce{GaV4S8}, polar domains in which ferroelastic strain is the primary order parameter can be observed.\cite{butykai2017}
The ferroic properties of halide perovskites seem characteristic of a {\it ferroelastic electret}.
A ferroelastic electret can be defined as a dielectric in which the primary order parameter, the spontaneous strain, is coupled to a quasi-permanent polarization.
Similar twin domains have also been observed in the tetragonal phase of $\ce{(FA_{0.85}MA_{0.15})Pb(Br_{0.85}I_{0.15})3}$, and $\ce{Cs_{0.05}(FA_{0.85}MA_{0.15})_{0.95}Pb(Br_{0.85}I_{0.15})3}$ thin film samples.\cite{Gratia2016,Tan2017}
%
%
Notably, the standard annealing procedure at ca. 100$^{\circ}$C means that all \ce{MAPbI3} thin-films will be subject to this phase transition. 
However, it has been reported that alloying \ce{MAPbI3} with FA$^+$ based halide perovskites reduces the temperature of the cubic-tetragonal phase transition.\cite{Weber2016}
%
%
Therefore, we do not expect that these mixed cation compounds will exhibit the same strain-induced deformation at room temperature.
%
%

%
%
%
%
%
%
%
%
%
\subsection{Imaging of ferroic domain behavior}
The nature of a ferroelastic state may be directly probed by the measurement of elastic hysteresis cycles.\cite{Tagantsev2010,Prasad1977}
Such measurements do not feature in the literature of halide perovskites, however, as the technique is  practically challenging.
The majority of experimental evidence instead arrives from the PFM data maps previously discussed.
PFM is a complex technique itself as measurements are sensitive to the effect of electrostatic, ionic and topological artifacts, which makes unambiguous interpretation difficult.\cite{Balke2015}
Consequently, characterization of the ferroelastic state and its coupling to a polar response has been complicated by conflicting results.
For example, a number of vertical PFM (v-PFM) measurements show lamellar domain contrast between high- and low-response piezoelectric regions -- as seen in Figures 3(a \& c) -- which suggest the existence of multiple response mechanisms.\cite{Hermes2016,Strelcove2017,Wei2018}
%
%
In support of this, Huang et al. report that electrostatic interactions or ionic activity are responsible for the low piezo-response measured in alternate domains in v-PFM data maps.\cite{Huang2018,Chen2014,Li2015}
The authors spatially correlate these low response regions with domains in lateral PFM measurements which show no piezo-response and subsequently argue that domain boundaries exist between a polar and a non-polar space group.\cite{Huang2018}
%
%

%
%
Reports of a consistently non-zero lateral PFM signal and of variation in the piezoresponse being restricted to domain boundaries -- as shown in Figure 5 -- seem to contradict this behavior.\cite{Rohm2017,Vorpahl2018,Kutes2014}
%
%
Vorpahl et al. suggest instead that inhomogeneity in the PFM amplitude is due to depolarization fields.
The authors further argue that their lateral PFM measurement, observed in Figure 5(c), provides evidence that the polarization lies parallel to the surface and that PFM phase contrast, observed in Figure 5(d), emerges due to a $\sim180\degree$ offset in the orientation of electrostatic dipoles.\cite{Vorpahl2018}
Polar domain formation is sensitive to the crystal grain size, morphology, and quality.\cite{lines1977}
%
%
Strain can lower the energetic barrier to ionic bond dissociation such that the density of point defects may be significantly enhanced within and between domains.\cite{Walsh2018,Jones2018}
%
This effect can explain twin domain contrast observed in \ce{CH3NH3+} chemical composition data maps.\cite{Liu2018}
%
%
We suggest that the variation in observed behavior can be linked to differences in initial strain distribution (e.g. due to stoichiometry or substrate effects) and the subsequent response of the crystal. 
\section{Outlook}
Hybrid halide perovskites -- and organic-inorganic crystals in general -- are examples of complex dielectric materials that feature a range of polarization mechanisms.
These include the perturbation of the electron clouds around atomic centers, the displacement and rotation of ions and molecules, as well as the transport of charged species (ions and electrons).
Based on a survey of the literature for \ce{CH3NH3PbI3} we recommend values of
$\epsilon_{optic} = 5.5$ for high-frequency processes, $\epsilon_{ionic} = 16.5$ for the ionic contribution, which yields a bulk dynamic dielectric response, of $\epsilon_{r} = 22.0$.
A rigorous description of even a homogeneous bulk material is challenging and requires a combination of theoretical (statistical mechanical) and experimental techniques.
In his 1949 monograph \textit{Theory of Dielectrics}, Fr\"{o}hlich noted: ``That this application is far from trivial is shown by some of the controversies in the literature''. 
This is certainly true for the halide perovskites.
However, from an assessment of the current literature, we conclude that the intrinsic bulk dielectric response is dominated by the displacement of ions as described the phonon modes of the crystal. 
The ferroic properties of \ce{MAPbI3} and related materials have attracted significant attention over the past five years.
It is difficult to separate lattice polarization from charge transport and surface effects.
We highlight the growing evidence supporting the assignment of room-temperature domain structures as ferroelastic, arising from the cubic-to-tetragonal phase transition that occurs during the standard annealing process of thin-films. 
Such domains can be associated with trapped charged defects and charge carriers.
We conclude that halide perovskites can be classified as ferroelastic electrets, which raises many exciting possibilities for engineering their polarization states and lifetimes. 
A significant amount of work remains in physical characterization of these materials and the development of quantitative models to describe the full range of physical processes at play in operating photovoltaic devices.   
\begin{acknowledgments}
We thank P. R. F. Barnes, L. Herz, S. Stranks and R. W. Whatmore 
for useful discussion on polarization, piezoresponse and perovskites.
This research has been funded by the EPSRC (Grant No. EP/K016288/1).
AW is supported by a Royal Society University Research Fellowship.
We are grateful to the UK Materials and Molecular Modelling Hub for computational resources, which is partially funded by EPSRC (EP/P020194/1).
\end{acknowledgments}
\bibliographystyle{apsrev4-1}
\bibliography{Bibliography.bib}
\end{document}